\documentclass[fleqn,usenatbib]{mnras}

\usepackage{newtxtext,newtxmath}
\usepackage{siunitx}

\usepackage[T1]{fontenc}
\usepackage{ae,aecompl}

\DeclareRobustCommand{\VAN}[3]{#2}
\let\VANthebibliography\thebibliography
\def\thebibliography{\DeclareRobustCommand{\VAN}[3]{##3}\VANthebibliography}

\usepackage{graphicx}	
\usepackage{amsmath}	
\usepackage{amsmath,amstext}
\usepackage{gensymb}
\usepackage[normalem]{ulem}

\usepackage{color} 
\usepackage{hyperref} 
\usepackage{makecell} 

\include{newCommands} 

\def\Msun{\hbox{M$_{\astrosun}$}}             
\def\Rsun{\hbox{R$_{\astrosun}$}}

\def\Mearth{\hbox{M$_{\oplus}$}}
\def\Rearth{\hbox{R$_{\oplus}$}}
\def\degr{\hbox{$^\circ$}}
\def\teff{T$_{\rm eff}$}
\def\logg{log~{\it g}}
\def\met{[Fe/H]}

\def\Msun{\hbox{$\mathrm{M}_{\odot}$}}       
\def\Rsun{\hbox{$\mathrm{R}_{\odot}$}}

\def\degr{\hbox{$^\circ$}}

\title[A highly mutually-inclined, compact warm-Jupiter system KOI-984 ?]{A highly mutually-inclined, compact warm-Jupiter system KOI-984 ?}

\author[L. Sun et al.]{L. Sun,$^{1,2}$
P.~Ioannidis,$^{3}$
S.~Gu,$^{1,2,4}$
J.~H.~M.~M.~Schmitt,$^{3}$
X.~Wang,$^{1,2,4}$
M.~B.~N.~Kouwenhoven,$^{5}$
\newauthor
V.~Perdelwitz,$^{6,3}$
F.~Flammini~Dotti$^{5}$
and S.~Czesla$^{3}$
\\
$^{1}$Yunnan Observatories, Chinese Academy of Sciences, Kunming 650216, China\\
$^{2}$Key Laboratory for the Structure and Evolution of Celestial Objects, Chinese Academy of Sciences, Kunming 650216, China\\
$^{3}$Hamburger Sternwarte, Universität Hamburg, Gojenbergsweg 112, 21029 Hamburg, Germany\\
$^{4}$ School of Astronomy and Space Science, University of Chinese Academy of Sciences, Beijing 101408, China\\
$^{5}$Department of Physics, School of Science, Xian Jiaotong-Liverpool University (XJTLU), 111 Renai Rd., Suzhou Dushu Lake Science and Education\\
 Innovation District, Suzhou Industrial Park, Suzhou 215123, China\\
$^{6}$Department of Physics, Ariel University, Ariel 40700, Israel
}
\date{Accepted on 08/11/2021 by MNRAS}

\pubyear{2021}

\begin{document}
\label{firstpage}
\pagerange{\pageref{firstpage}--\pageref{lastpage}}
\maketitle

\begin{abstract}
The discovery of a population of close-orbiting giant planets ($\le$ 1 au) has raised a number of questions about their origins and dynamical histories. These issues have still not yet been fully resolved, despite over 20 years of exoplanet detections and a large number of discovered exoplanets. In particular, it is unclear whether warm Jupiters (WJs) form in situ, or whether they migrate from further outside and are even currently migrating to form hot Jupiters (HJs). Here, we report the possible discovery and characterization of the planets in a highly mutually-inclined ($I_{\rm mut}\simeq 45 \degr$), compact two-planet system (KOI-984), in which the newly discovered warm Jupiter KOI-984$c$ is on a 21.5-day, moderately eccentric ($e\simeq 0.4$) orbit, in addition to a previously known 4.3-day planet candidate KOI-984$b$. Meanwhile, the orbital configuration of a moderately inclined ($I_{\rm mut}\simeq 15 \degr$), low-mass ($m_{c}\simeq 24 \Mearth$;$P_b\simeq 8.6$ days) perturbing planet near 1:2 mean motion resonace with KOI-984$b$ could also well reproduce observed transit timing variations and transit duration variations of KOI-984$b$. Such an eccentric WJ with a close-in sibling would pose a challenge to the proposed formation and migration mechanisms of WJs, if the first scenario is supported with more evidences in near future; this system with several other well-measured inclined WJ systems (e.g., Kepler-419 and Kepler-108) may provide additional clues for the origin and dynamical histories of WJs.

\end{abstract}




\section{Introduction}\label{sec:introduction}

\par With the success of the {\it Kepler} mission, exoplanetary science has entered a new era~\citep{Borucki2010}. Among the large number of exoplanets discovered, many have orbital and physical properties that are very different from the planets in our own Solar system. This is, for example, the case for hot and warm Jupiters, which are Jupiter-sized planets with orbital periods of $<10$ d and with orbital periods between 10 and $~200$~d, respectively. Our understanding of the formation of these classes of exoplanets is still limited~\citep{Dawson2014,Dawson2018,Winn2015}. For Hot Jupiters (HJs; semi-major axes $\le 0.1$~au), the general consensus has emerged that they cannot form at their present locations, and must have migrated from further outside through disc migration and/or high-eccentricity migration~\citep{Lin1996,Kley2012,Baruteau2014,Boley2016,Dawson2018}. However, the origin of Warm Jupiters (WJs, semi-major axes between 0.1 and 1~au) remains an unsolved problem. 

\par There is evidence showing the existence of two distinct populations of warm Jupiters. The majority of these planets are characterized by low eccentricities ($e<0.2$), nearby super-Earth companions commonly in nearly-coplanar orbits, and a dearth of external Jovian-mass companions~\citep{Dong2014,Huang2016,Barragan2018,Hjorth2019,Anderson2020}. The remaining warm Jupiters are characterized by moderately eccentric ($e>0.4$) orbits, often accompanied by external Jovian-mass companions which are mutually inclined and apsidally misaligned~\citep{Dawson2014,Masuda2017}. Studying the eccentricity and companionship of warm Jupiter systems, researchers have proposed that warm Jupiters originate from two different formation paths: high-eccentricity migration and in situ formation~\citep{Dong2014,Huang2016,Boley2016}. When WJs form through high-eccentricity migration, they are expected to have experienced secular eccentricity oscillations with the aid of outer close-by high-mass companions, and thus have high eccentricities ($e>0.4$) and no low-mass inner companions~\citep{Dong2014,Mustill2015,Mustill2017,Anderson2017}; while if they form in situ, they should have small eccentricities ($e<0.2$) and inner low-mass siblings with small mutual inclinations ~\citep{Huang2016,Petrovich2016,Boley2016}. 

\par WJs are close enough to their host stars that they likely have experienced significant migration, but distant enough from their hosts that tidal effects are unlikely to have erased the potential imprints of their migration histories~\citep{Li2016,Dawson2018}. Therefore, the detection and characterization of WJs play a key role in figuring out the origins and dynamical histories of close-orbiting giant planets. However, the population of known warm WJs around nearby stars that are available to detailed characterization is still very small. 

\par In this paper, we present the dynamical modeling of Transit Timing Variations (TTVs) and Transit Duration Variations (TDVs) of KOI-984$b$. The paper is structured as follows. In Section \ref{koi-984}, we describe the fundamental properities of this system. Section \ref{data analysis} outlines the TTVs and newly measured TDVs of KOI-984$b$ used for our analysis. Then we give our dynamical modeling and long-term stability analyses in Section \ref{dynamical modeling} and Section \ref{dynamical stability}, respectively. The results are discussed in Section \ref{discussion}. Finally, in Section \ref{conclusions} we summarize this work.

\section{The KOI-984 planetary system}\label{koi-984}

\par The star KOI-984 was observed by the {\it Kepler} space telescope during its primary mission~\citep{Borucki2010}, for a time span of $\sim 1470$ days (Q1-17 of the {\it Kepler} data). The data revealed a transiting candidate KOI-984.01 (hereafter: KOI-984$b$)~\citep{Ford2012,Batalha2013}. ~\citet{Law2014} reported that  KOI-984$b$ had an orbital period of about 4.29~days and a radius of 4.5 $\Rearth$, which shows that the planet is likely located in the hot Neptunian desert~\citep{Szabo2011,Mazeh2016}. Furthermore, KOI-984$b$'s transit times exist a strong TTVs with an amplitude of roughly 4-hours~\citep{Mazeh2013,Holczer2016}.

\par KOI-984 is a late-G-type dwarf star located at a distance of $229.63\pm1.24$~pc~\citep{Gaia2018,Lindegren2018,Stassun2018}. It has an effective temperature of \teff = 5295 $\pm$ 150~K and an iron abundance of \met = 0.12 $\pm$ 0.1 dex~\citep{Deck2015}(see Table~\ref{table1} for further detail). KOI-984 has a wide-orbit stellar companion at an angular separation of $(1.8\pm 0.06)''$~\citep{Law2014}, which is slightly brighter than KOI-984 by $\Delta K_{p}= 0.072 \pm 0.026$ mag~\citep{Deck2015}. Photometric modulations with an amplitude of about 2$\sim$4\% and a period of $7.98 \pm 0.01$ days were clearly seen in the mixed {\it Kepler} light curve of KOI-984 and the close stellar companion~\citep{Morton2014}. Below, we treat the mixed light curves as those of KOI-984 alone, since we know the differential magnitude, $\Delta K_p$, in the {\it Kepler} band, and therefore the flux contribution from the stellar companion, can be well accounted for with a dilution factor.

\par KOI-984 was spectroscopically observed once with the Keck/HIRES spectrograph~\citep{Vogt1994}. Furthermore, \cite{Deck2015} observed two more transits of KOI-984$b$ by using the 48-inch telescope at Fred Lawrence Whipple Observatory. The transit times were acquired by them through modeling these two low signal-to-noise light curves\footnote{When we almost independently completed our research of this system, we occasionally found the on-line material about the TTV research of KOI-984 system carried out by~\cite{Deck2015} (\url{https://dspace.mit.edu/handle/1721.1/91078}). Our conclusion is totally different from theirs, however, because of their much narrower search space of KOI-984$c$'s orbital parameters.}.

\section{Data analysis}\label{data analysis}

\subsection{Data preparation and transit searching}

The {\it Kepler} data of KOI-984 were retrieved from the MAST archive\footnote{\url{https://archive.stsci.edu/kepler}}, which recorded the 17 quarters of {\it Kepler} photometry; we use both long- and short-cadence data  (only available for quarters Q9 to Q17). The PDC\_SAP format data are employed for our analysis. One easily find peak-to-peak variations of $2\sim$4\% in the light curve, indicating that at least one of the host star and the close companion is relatively active, with a clear rotation period of about 8 days. 

To characterize the system, it is vital to eliminate the influences from stellar activity as much as possible. We adopt the similar manner as in~\cite{Ioannidis2014} and~\cite{Sun2019} to extract transits from {\it Kepler} data and detrend each transit light curve. That is, we use a window of about triple the transit duration, approximately centered on the mid-transit time of each transit light curve, and detrend each light curve with a second-order polynomial fitted to the out-of-transit data. We select the transits only near the local maxima of KOI-984's light curves for the latter transit modeling (see Section \ref{transit modeling} for details),  in order to account for the possible influences of stellar activity on the measurements of transit parameters. However, the error of transit timing measurements induced by stellar activity is typically in several minuts~\citep{Ioannidis2016}, which might be neglected compared with over 4-hour TTVs induced by planetary gravitational interactions. To take advantage of the constraints on the dynamical modeling from all available transits, the transit times of KOI-984$b$ measured by~\cite{Holczer2016} were acquired from the literature and used in our TTV analysis.

We also downloaded TESS data of KOI-984 from the MAST archive~\citep{Ricker2015}, in order to search extra transit events of KOI-984$b$. However, we did not detect any convincing transit events due to the low signal-to-noise light curves of this target collected by TESS.

\subsection{Model fitting}\label{transit modeling}

Star spots do not only generate bumps in a transit light curve when they are occulted by a transiting planet, but it can also lead to variable depths of transit light curves normalized with the out-of-transit data, which originates from the variations of total flux of the host star due to the stellar magnetic activity~\citep{Czesla2009,Ioannidis2016,Sun2017}. 

\begin{figure*}
\centering
\includegraphics[width=\linewidth]{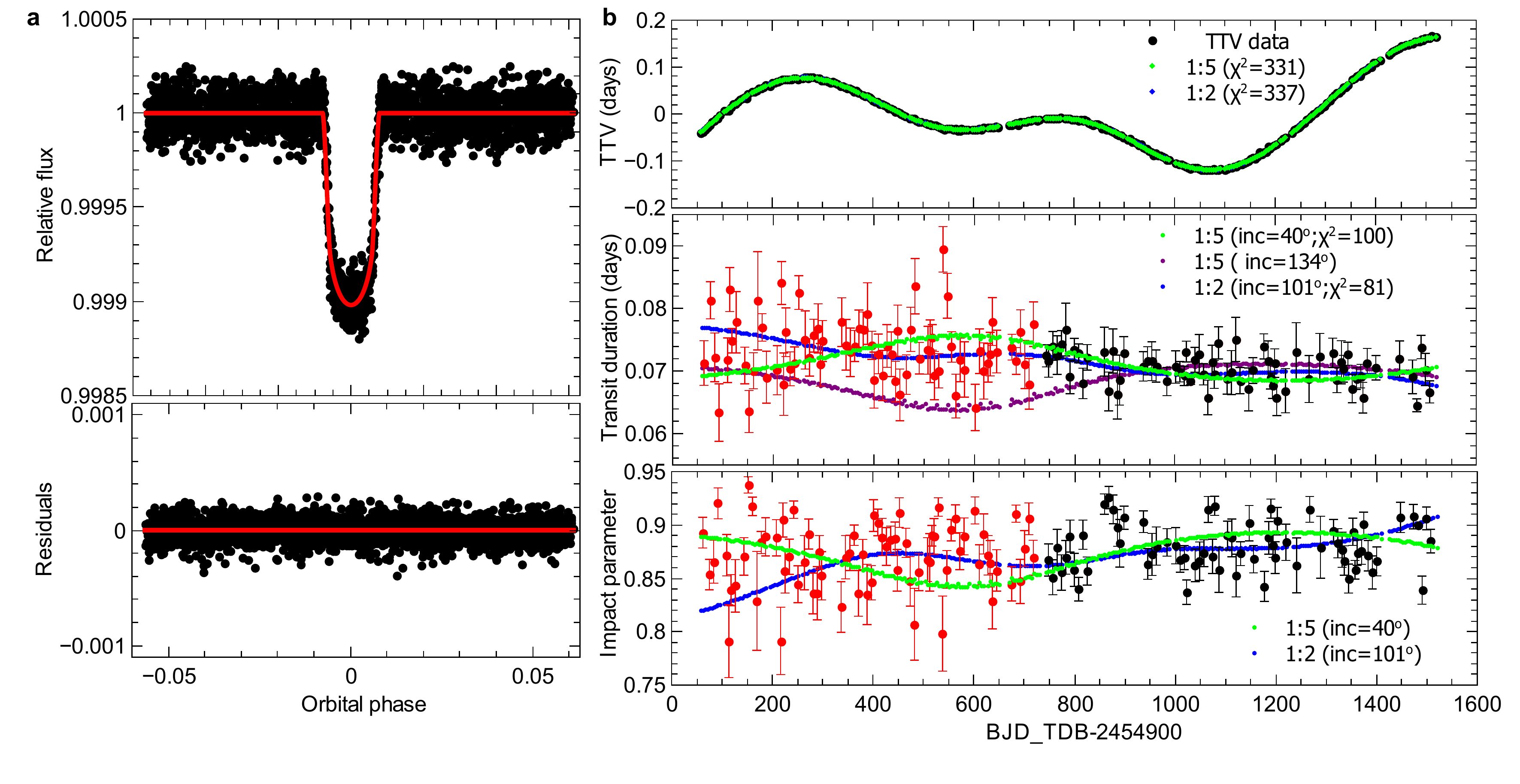}
\caption{KOI-984$b$'s transit light curves, TTVs, TDVs and impact parameters with the respective fitting models. \textbf{a}, Upper panel : filled circles are binned short-cadence photometry observations; the red curve represents the optimal fitting model. Lower panel: filled circles and red line are respectively residuals and reference, which represents the perfect matching between observed data and model. \textbf{b}, Upper panel : black scatters represent measured TTVs based on {\it Kepler} long-cadence photometry. Green scatters are the optimal fittings of the orbital architecture near 1:5 MMR with KOI-984 $b$, while blue scatters are those near 1:2 MMR. The fitting result near 1:5 MMR is superior to that near 1:2 MMR by $\Delta \chi^2=6$, though it seems that both fittings are very similar to one another. With the free degree of 295, the reduced $\chi^2_{\nu}$ is 1.12 for near 1:5 MMR orbital architecture and thus KOI-984$b$'s TTVs are not overfitted. Middle panel: red and black scatters are transit duration measurements of KOI-984$b$ based on {\it Kepler} long- and short-cadence photometry, respectively. The green and purple scatters represent the optimal fitting models of two different orbital architectures near 1:5 MMR for black scatters, while the blue scatters are those near 1:2 MMR. The fitting result of near 1:2 MMR is superior to that of near 1:5 MMR for black scatters by $\Delta \chi^2=19$. This mismatch to red scatters helps us rule out the orbital architecture with largest inclination. Lower panel: red and black scatters are impact parameter measurements of KOI-984$b$ based on {\it Kepler} long- and short-cadence photometry, respectively. The green and blue scatters represent  impact parameter models produced by the optimal fitting models of KOI-984$b$'s TTV and TDV based on the orbital architectures near 1:5 MMR and 1:2 MMR, respectively. }
\label{plot1}
\end{figure*}

In order to accurately measure the transit parameters, we use the Spot and Transit Modeling Tool (STMT) developed by \citet{Sun2017} to model the transit light curves of KOI-984. STMT is capibale of simultaneously modeling the joint effects of spots and planets on the light curve (see~\cite{Sun2017} for further details). The quadratic limb-darkening law is used during our transit light curve modeling; The limb darkening coefficients are derived through interpolation of the coefficient tables of~\cite{Claret2013} and fixed in the modeling (see Table~\ref{table2} for the adopted values). The prior of free parameters in transit modeling are also listed in Table~\ref{table2}. The transit light curves recorded in the short-cadence mode are utilized to calculate the transit parameters of KOI-984$b$. We only select the transit light curves approached to the local maximum of KOI-984's light curves to caculate the physical parameters, because these transit light curves are expected to be less contaminated by star spots. The short-cadence data could provide much better constraints for the transit parameters than the long-cadence data (i.e., only four points had been recored for each of KOI-984$b$'s transits in long-cadence mode). Moreover, when many short-cadence light curves with sufficient sampling are modeled simultaneously, the influence of stellar activity could be well alleviated on the measurements of transit parameters. According to the measurements of differential magnitude of KOI-984's stellar companion with KOI-984~\citep{Law2014,Deck2015}, we computed the dilution factor and fixed it in our transit modeling. The contaminated transit light curve is simply formulated as $\delta F^{'}=\Delta F/(F_0*(1+F1/F_0))$, where $\Delta F/ F_0 $ represents the normal transit light curve and $F1/F_0$ is the flux ratio of both stars in observed band (i.e., $F_0$ is the flux of the planet host). During the transit modeling, $F_1/F_0$ is fixed to 1.0683. Moreover, the empirical relation of~\citet{Enoch2010} between main-sequence exoplanet hosts' masses and their stellar atmospheric parameters (i.e., \teff, $\rho_{\star}$ and \met) are employed in the transit modeling to acquire KOI-984's mass and radius. For the transit modeling, we use MCMC to sample the posteriors of transit parameters for 60000 samples and obtain the statistics of the remaining samples to acquire the the posterior distributions after remove the first 20\% burn-in samples. We employ the Gelman-rubin statistics to ensure the convergence of the MCMC sampling. In addition, we employ the MCMC code to model transit light curves for additional ten times and check the consistence of derived parameters to test the convergence of our transit modelings. Finally, we find that both methods prove the convergence of the MCMC sampling. In Figure \ref{plot1}, we show our best-fit results of transit light curves; we further list the best-fit system parameters of the transiting planet in Table~\ref{table3} derived from modeling the transit light curves.  

Furthermore, in order to derive more reliable transit durations and impact parameters of KOI-984$b$, we individually fitted each of short- and long-cadence transit light curves that met the above selection criterion. Because these light curves were in principle less affected by brightness variation due to spots and spot crossing events , which were capable of influencing the measured transit parameters including the transit durations. During fitting each of selected transit light curves, the radii ratio between planet and host star was fixed to previous fitting result. In Figure~\ref{plot1}, we show our fitting results of durations and impact parameters of KOI-984$b$.

\section{Dynamical modeling}\label{dynamical modeling}

\subsection{Prior information of orbital parameters}

We utilize our well-tested TTV inversion code~\citep{Sun2019}, which is based on TTVFast~\citep{Deck2014} to calculate transit times, to inverse observed KOI-984$b$'s TTVs. It employs Genetic Evolution Markov Chain (GEMC)~\citep{Tregloan-Reed2013} and DiffeRential Evolution Adaptive Metropolis (DREAM) algorithms~\citep{Laloy2012,Sun2017,Sun2019} to carry out parameter optimization and estimation. Recently, in order to improve the efficiency of its parameter optimization, we incorporate the Levenberg-Marquardt (LM) algorithm in the code to acquire a series of initial guesses of GEMC chains, which enhances the speed to converge to the global optimal solution because of the proximity of LM's initial guess to the local optimization solution. In addtion, we build another TTV modeling code through combining TTVFast and the multimodal nested sampling routine MultiNest to independently analyze the TTVs~\citep{Feroz2009,Feroz2019,Buchner2014}. The MultiNest is specially designed to compute the Bayesian evidence in complex and multimodal parameter space in efficient manner. Although the primary goal of MultiNest is to compute the Bayesian evidence, it also produces the posterior parameter distribution as a by-product. Marginalizing these posteriors allows us to acquire the nominal parameter estimation and the associated uncertainties\citep{Feroz2009}. We hereafter label the two TTV inversion codes as MCMC-based code and MultiNest-based one, respectively.

As mentioned in Section \ref{koi-984}, the only planet candidate that was previously reported, KOI-984$b$, shows large TTVs with an amplitude of over 4~hours. This implies that there is (at least) one more unseen planet in the orbit around KOI-984; the additional planet(s) may be in or near a mean-motion resonances (MMRs) state with KOI-984$b$, in which the TTV signals induced by planetary gravitational interactions are amplified~\citep{Agol2005,Holman2005,Holman2010,Nesvorny2012,Jontof2015,Sun2019}. We first follow a similar methodology as in the literature~\citep{Nesvorny2012,Sun2019} by utilizing our MCMC-based code to search for the optimal orbital architecture of the perturber(s).  

TTV patterns induced by planetary gravitational interactions typically mostly depend on the mass and orbital period of the perturber and both interacting planets' orbital eccentricities and even argument of pericenters~\citep{Agol2005,Holman2005,Nesvorny2009,Nesvorny2012,Xie2014}. Therefore, we set apropriate priors to these parameters during the first-step search: 
(i) For the mass of the perturber, we impose a uniform prior between $10^{-8}M_*$ and $0.009M_*$. 
(ii) In order to cover possible architectures, the search space of the perturber's orbital period is set uniformly between 1 and 65 days: the lower limit corresponds to approximately one fourth of KOI-984$b$'s orbital period, while the upper limit is set to fifteen times of KOI-984$b$'s orbital period. We split the search space of the perturber's orbital period into multiples of 0.5-day intervals for $P_c > P_b$, while split that into multiples of 0.2-day intervals for $P_c < P_b$. The perturber's orbital period is treated as free parameters in each interval during our modeling of the observed TTVs. 
(iii) The eccentricity prior of KOI-984$b$ is uniformly imposed between $e=0$ and $e=0.5$, as required by the long-term stability criterion of the system, which sufficiently covered the eccentricity range of {\it Kepler} compact multi-planet systems~\citep{Lissauer2011,Fabrycky2014,Xie2016}. The eccentricity prior of the outer perturber (i.e., $P_c > P_b$) is uniformly set between $e=0.0$ and $e=1.0$, but the inner perturber (i.e., $P_c < P_b$) shares an identical eccentricity prior with KOI-984$b$. During the TTV modeling, we set the combination of eccentricity and argument of pericenter ($\sqrt{e}\cos\omega$,$\sqrt{e}\sin\omega$) as free parameters to ensure uniform sampling of eccentricity, where $\omega$ denotes the argument of pericenter~\citep{Eastman2013}. (iii) Moreover, the mutual inclination also weakly affects the TTV pattern, as compared with the above-mentioned parameters. Therefore, we impose a uniform prior between  0\degr and 30\degr on the mutual inclination ~\citep{Nesvorny2009}. The mutual inclination is determined by two planets' inclinations (i.e., $i_b$ and $i_c$) and the differential longitude of ascending nodes, $\Delta \Omega$.  KOI-984$b$'s inclination is well-constrained via modeling of the transit light curves, so we explicitly fix KOI-984$b$'s inclination to the fitting result. To simplify the differential longitude of ascending nodes, we set $\Omega_b$=0\degr and thus $\Omega_c$=$\Delta \Omega$. Therefore, $i_c$ and $\Omega_c$ are uniformly sampled on the intervals U($i_b$-30\degr,$i_b$+30\degr) and U(-30\degr,30\degr). 

Subsequently, we use the MultiNest-based code with much wider mutual inclination prior than that of first step search, aiming to refine the local optimum solutions derived by MCMC-based code, but the remaining prior distributions are kept same with the first-step search. For the MultiNest-based code search, $i_c$ and $\Omega_c$ are uniformly sampled on the intervals U($i_b$-60\degr, $i_b$+60\degr) and U(-60\degr,60\degr) respectively, which is helpful to examine whether larger mutual inclinations can provide better fittings to KOI-984$b$'s TTV. Also see table~\ref{table1} for further information.     

\subsection{Grid search on two-planet assumption}

\begin{figure*}
\centering
\includegraphics[width=\linewidth]{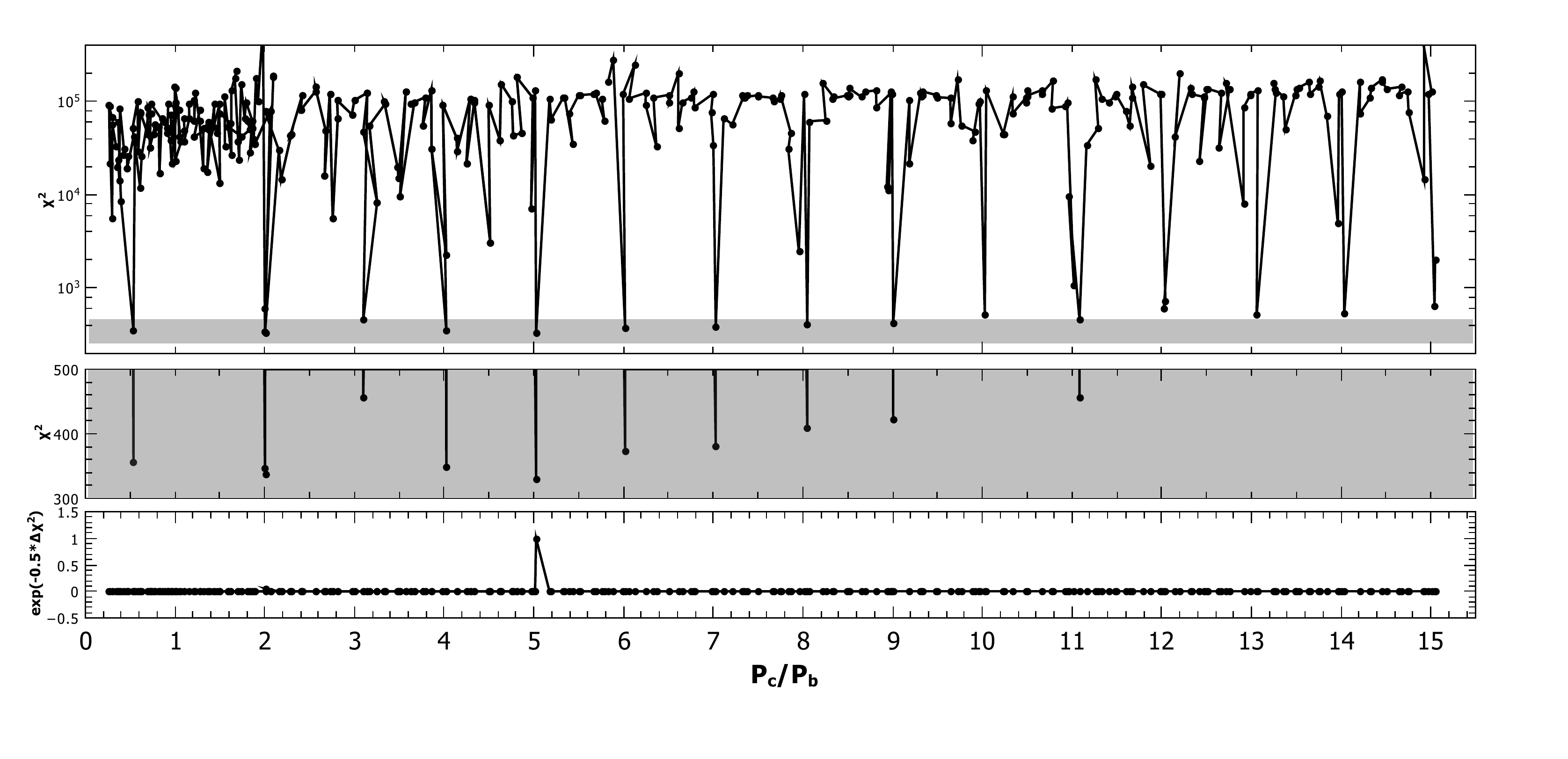}
\caption{The $\chi^2$ of two-planet model TTV fittings versus the period-ratio of the purterber and KOI-984$b$ in upper panel. The gray region in upper panel is zoomed in the middle panel. For clearly visualizing and comparing all local optimum solutions, their likelihoods obtained in initial search are normalized with that of the global optimal solution near 1:5 MMR and depicted in lower panel.}
\label{plot2}
\end{figure*}

We first test the two-planet architecture. That is, the observed TTVs of KOI-984$b$ is completely induced by an unseen planet KOI-984$c$. We search the parameter space using the TTV inversion code and record the minimum $\chi^2$ in each orbital interval. We find that the $\chi^2$ dramatically drops when the trial period is near the integer multiples of KOI-984$b$'s orbital period, that is, these local optimum solutions were near MMRs (i.e., $P_b/P_c \simeq $1:2, 1:3, 1:4, 1:5, 3:5, 2:3, 3:2, 5:3, 2:1, etc.; hereafter $m:n$ denotes $P_b/P_c \simeq m:n$ for convenience). Subsequently, we iteratively model KOI-984$b$'s TTVs around these local optimum solutions for several times, in order to refine these solution. Secondly, we employ the nested sampling tool MultiNest~\citep{Feroz2009,Feroz2019,Buchner2014} to search the parameter spaces around those local optimum solutions, to check whether the orbital architectures with larger mutual inclinations can fit the KOI-984$b$'s TTVs better. To obtain reliable parameter estimation, we employed 1000 live points in Multinest-based code to mapping these parameter spaces and terminated it when evidence precision reached below 0.2. Although the dynamical modeling with MultiNest are much more time-consuming than our previous searches, the $\chi^2$ of most of optimum solutions improved significantly. Finally, four lowest $\chi^2$, near the 2:1, 1:2, 1:4 and 1:5 MMRs are 357, 337, 355 and 331, respectively (see Figure \ref{plot2} for more details.) Note that the total number of TTV data points and free parameters are 308 and 13, respectively. The $\chi^2$ of other local optimum solutions, however, are much larger than 370.

\par In addition, there is a pair of optimal solutions near each MMR being able to fit the TTV with identical $\chi^2$, in which one of inclinations is larger than 90\degr, and another is smaller than 90\degr. However, these two solutions are usually expected to generate different TDV patterns. In Figure \ref{plot1}, we display the TDVs generated by two optimal TTV solutions near the 1:5 MMR with different inclinations; it is apparent that the model with low inclinatin (i.e., $i_c=40\deg$) is more consistent with the measured TDVs than the one with large inclinaton. Furthermore, in order to examine the retrograde orbital architecture, we also searched the parameter space near the 2:1, 1:2, 1:4 and 1:5 MMRs by replacing previous prior of $\Omega_c$ with $U(150\degr, 210\degr)$. But we found that the $\chi^2$ of retrograde orbital architectures were much larger than those of prograde ones.  

\par
\begin{figure*}
\centering
\includegraphics[width=\linewidth]{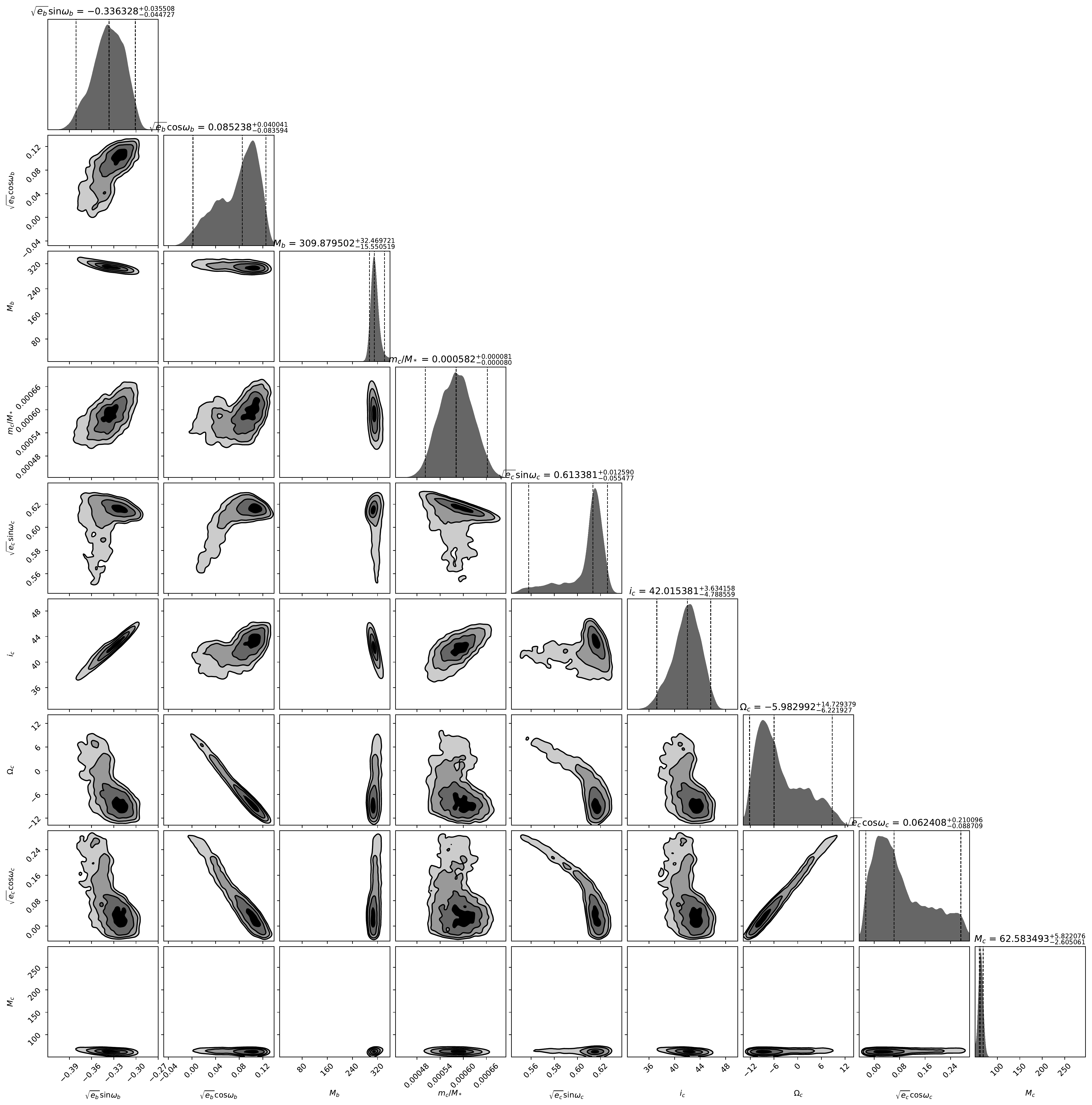}
\caption{The marginal distributions and pairwise correlations of free paramters associated with both planetary masses and orbital elements, which are derived from modeling KOI-984$b$' TTV using Multinest-based code based on the orbital architecture near 1:5 MMR. }
\label{plot3}
\end{figure*}

\begin{figure*}
\centering
\includegraphics[width=\linewidth]{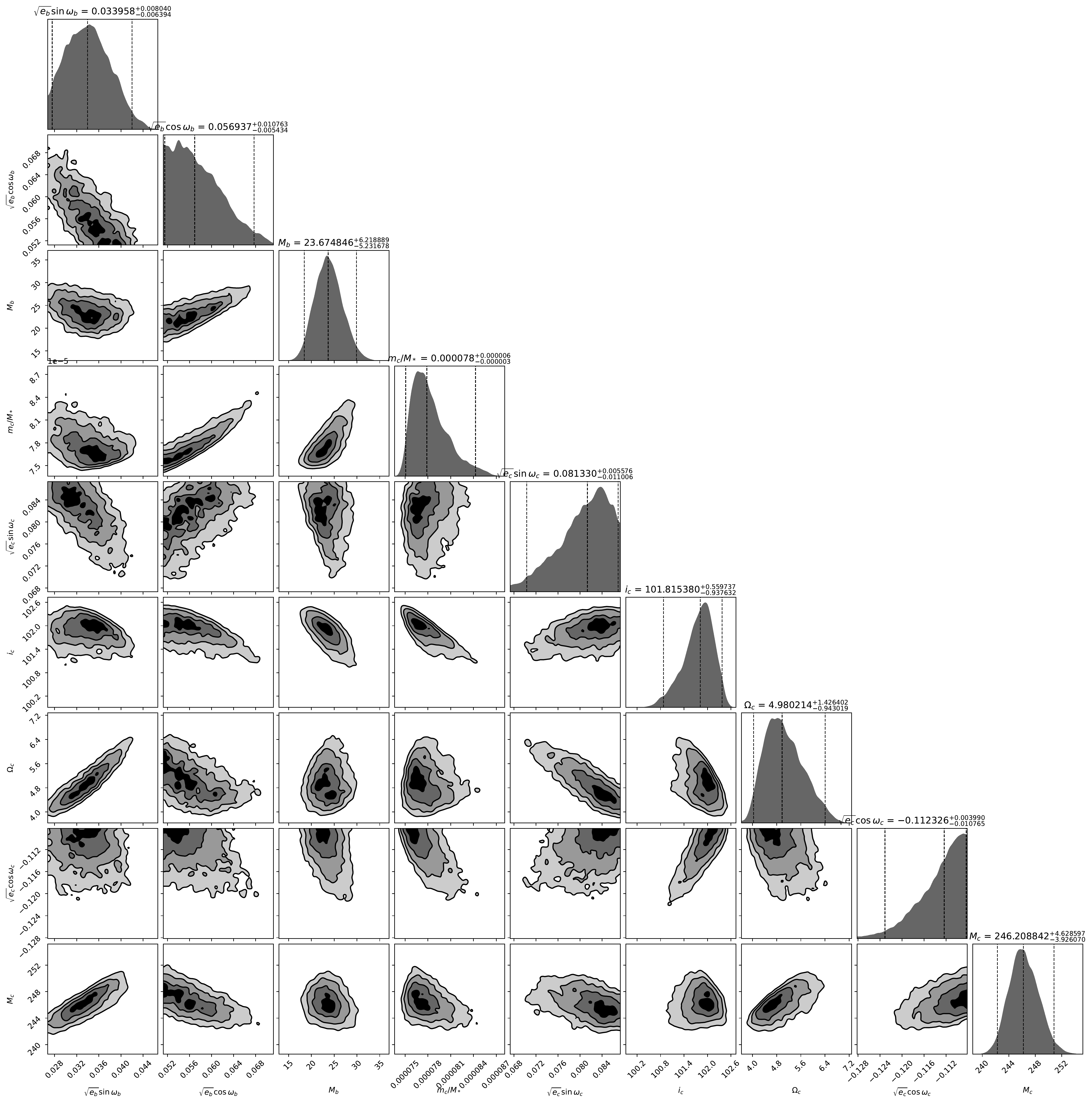}
\caption{The marginal distributions and pairwise correlations of free paramters associated with both planetary masses and orbital elements, which are derived from modeling KOI-984$b$' TTV using Multinest-based code based on the orbital architecture near 1:2 MMR. }
\label{plot4}
\end{figure*}

 Under the condition of $P_c\le 4P_b$, our search results are well consistent with those of \cite{Deck2015},  which implies that our search algorithm is efficient for inverting the TTV signal of KOI-984$b$. 
With the free degree of 295, the reduced $\chi^2_{\nu}$ of near 1:5 MMR is 1.12, which implies TTVs of KOI-984$b$ are not overfitted. The global optimal solution of TTV inversion is a warm-Jupiter near 1:5 MMR with KOI-984$b$. 
For the slightly less optimal ($\Delta \chi^2=7$) solution near 1:2 MMR, the perturber is a hot, Neptune-mass ($m_c\simeq24\Mearth$) planet. Athough the reduced $\chi^2_{\nu}$ of the orbital configuration near 1:5 MMR is slightly 
larger than unity, it is consistent with the expected reduced $\chi^2_{\nu}=1\pm 2\sqrt{1.0/295}$ at $1\sigma$ confidence. Based on either the inclination of 40\degr  (or 134\degr) near 1:5 MMR or 69\degr (101\degr) near 1:2 MMR, 
KOI-984$c$ is a non-transiting planet, which is consistent with the transit signals of only one planet detected in {\it Kepler} photometry. The optimal model TTVs are presented in Figure~\ref{plot1} with the measured TTV signal, as well 
the model TTVs near 1:2 MMR. In Figures~\ref{plot3} and \ref{plot4} , we show the marginal distributions and pairwise correlations of free paramters associated with both planets' masses and orbital elements, which are derived from 
the inversion of KOI-984$b$' TTV by using Multinest-based code.

\begin{figure*}
\centering
\includegraphics[width=\linewidth]{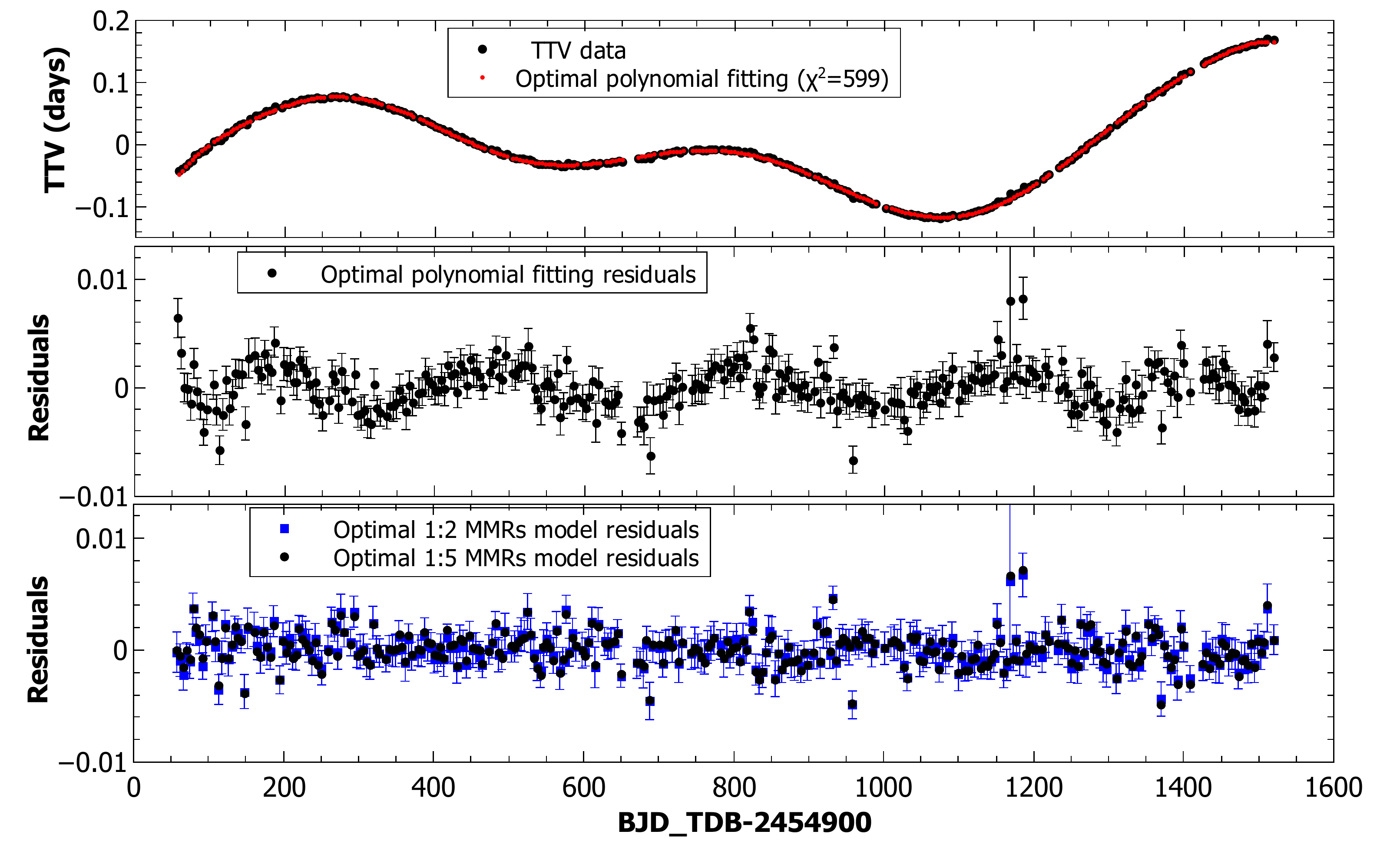}
\caption{A 10th order of polynomial is fitted and removed from the TTV data, and the resulting residuals are demonstrated in upper panel and middle one, respectively. The chopping variations (i.e. oscillation signals in the residuals of middle panel) are clearly detected for KOI-984$b$, which contributes to the discovery of unique solution of the perturber's orbital period and the high precision of the measurement of its mass. In addition, we show the fitting residuals of optimal 1:2 and 1:5 MMR models to the TTVs in lower panel, to help visualize different models' fitting residuals. }
\label{plot5}
\end{figure*}

We generate a second figure following the manner of~\citet{Agol2021} in which a polynomial (i.e., with an order 10) is fitted and removed from the TTV data, and the resulting difference is demonstrated in Figure \ref{plot5}. The result shows short-timescale variations that are (partially) associated with the synodic periods of both adjacent planets, typically referred to as “chopping.” The chopping signals not only reveal the perturbing planet's orbital period, but also encode the mass ratios of the companion planets to the star without the influence of the eccentricities and thus provide a constraint on the planet–star mass ratios which are less influenced by degeneracies with the orbital elements \citep{Lithwick2012,Nesvorny2014,Deck2015a}. The chopping variations are clearly detected for KOI-984$b$, which contributes to the discovery of unique solution of the perturber's orbital period and the high precision of the measurement of its mass. In addition, we show the fitting residuals of optimal 1:2 and 1:5 MMR models to the TTVs in Figure \ref{plot5}, to help visualize different models' fitting residuals.

\subsection{Grid search on three-planet assumption}\label{TTV modeling}

\par Considering the optimal solution of KOI-984's two-planet model is quite different from the majority of {\it Kepler} multi-planet systems, which are characterized by small mutual inclinations and low eccentricities~\citep{Lissauer2011,Fabrycky2014,Winn2015,Xie2016}, we also test the hypothesis that KOI-984$b$'s TTV signal is caused by two perturbing planets.

\par The TTV inversion of the three-planet architecture was carried out with a similar procedure as that used for the two-planet model. The only main difference is that the search of orbital periods of the perturbers is now two-dimensional (2D), which is more complicated as compared to the two-planet model.  For each search, the orbital periods of the two perturbers were restricted to 2D intervals of 0.5 (days) $\times$ 0.5 (days). We then search for the local minimum $\chi^2$ in each subspace. 

\par We find that the global minimum $\chi^2$ was 335, which was very close to the minimum $\chi^2$ of 331 of two-planet model, when the inner perturber is near the 1:1 MMR (namely, co-orbital architecture) with KOI-984$b$ and the outer one is near 1:6 MMR with KOI-984$b$ (i.e., $P_c=4.21$ days; $P_d=25.79$ days). In addition, other local $\chi^2$ minima are larger than 470, and thus this co-orbital architecture is significant. Co-orbital architecture is a by-product of some planetary formation and evolution models (e.g.,~\citet{Laughlin2002}); for example, in our Solar system there exist numerous trojans of Jupiter and Neptune. However, there is thus far no evident co-orbital architecture detected in exoplanetary systems~\citep{Leleu2019}.  

\par Occam's Razor suggests that the two-planet architecture is more credible compared to the co-orbital configuration, under nearly equivalent fitting degrees of KOI-984$b$'s TTV for two different models. In addition, the co-orbital configuration became unstable in a short time during our long-term orbital integration. Therefore, the highly-inclined, compact warm-Jupiter system is more reliable architecture of the perturber.

\subsection{Joint TTV and TDV modeling}\label{Joint ttv_tdv}

\par Based on previous analysis, an additional planet is needed to reproduce the measured TTVs of KOI-984 $b$. In this part, we jointly model the TTVs and TDVs of KOI-984 $b$, in order to test the optimal architecture derived by inverting the TTV only. Furthermore, the optimal solutions of TTV inversion are actually related to two different orbital architectures, which possess two totally different orbital inclinations, we expect that the TDVs of KOI-984$b$ could likely rule out one configuration. 

\par We jointly modeled all TTV data from the {\it Kepler} long-cadence photometry and two ground-based observation campaigns in the literature~\citep{Deck2015}, and the TDV data of pre-selected {\it Kepler} short-cadence photometry, since these pre-selected short-cadence data provide more reliable transit duration measurements. We initially perturbed the optimal TTV inversion solutions (i.e.,  $e_b$, $\omega_b$, $e_c$, $\omega_c$, $i_b$, $i_c$, and $\Omega_c$) to carry out the joint analysis, and the stellar radius was fixed to the value derived from previous transit modeling. It’s quite hard to converge for the joint TTV/TDV modeling, when both planets’ masses, eccentricities, arguments of the pericenters and mutual inclinations are treat as free parameters. So we fix the other parameters but the mutual inclinations to perform the joint TTV/TDV modeling, and find that the joint modeling does not significantly improve the constraining on the mutual inclination for KOI-984$b$ \& $c$ compared to the solution from TTV modeling only.  

\par Besides joint modeling of TTVs and TDVs on optimal solution near the 1:5 MMR, we carried out similar analysis on the solution near 1:2 MMR. We find that KOI-984$b$'s best fitting result to the TDVs for 1:2 MMR solution is superior than that for 1:5 MMR solution by $\Delta\chi^2$=19 (see Figure \ref{plot1}). However, we cannot explicitly conclude that the fitting result of 1:2 MMR for KOI-984$b$'s TTVs and TDVs is much better than that of 1:5 MMR, because the TDV measurements of KOI-984$b$ are much less accurate than TTV measurements. In addition, the TDV pattern generated by near 1:5 MMR architecture with $i_c$=134\degr does not match the TDVs only observed using long-cadence mode. 

\par Although we have measured the impact parameters in previous transit modeling, they was not included in the joint modeling of TTVs and TDVs. Because the stellar activity could affect the normalized transit light curves along with the variations of host star's brightness and thence bias the measurement of impact parameters, while (in principle) the measurement of transit durations cannot be influenced~\citep{Sun2017}. However, we still checked whether the measured impact parameters were consistent with the ones predicted by the joint modeling of TTVs and TDVs. It is clear that they are consitent with each other in Figure~\ref{plot1}.

\begin{figure*}
\centering
\includegraphics[width=\linewidth]{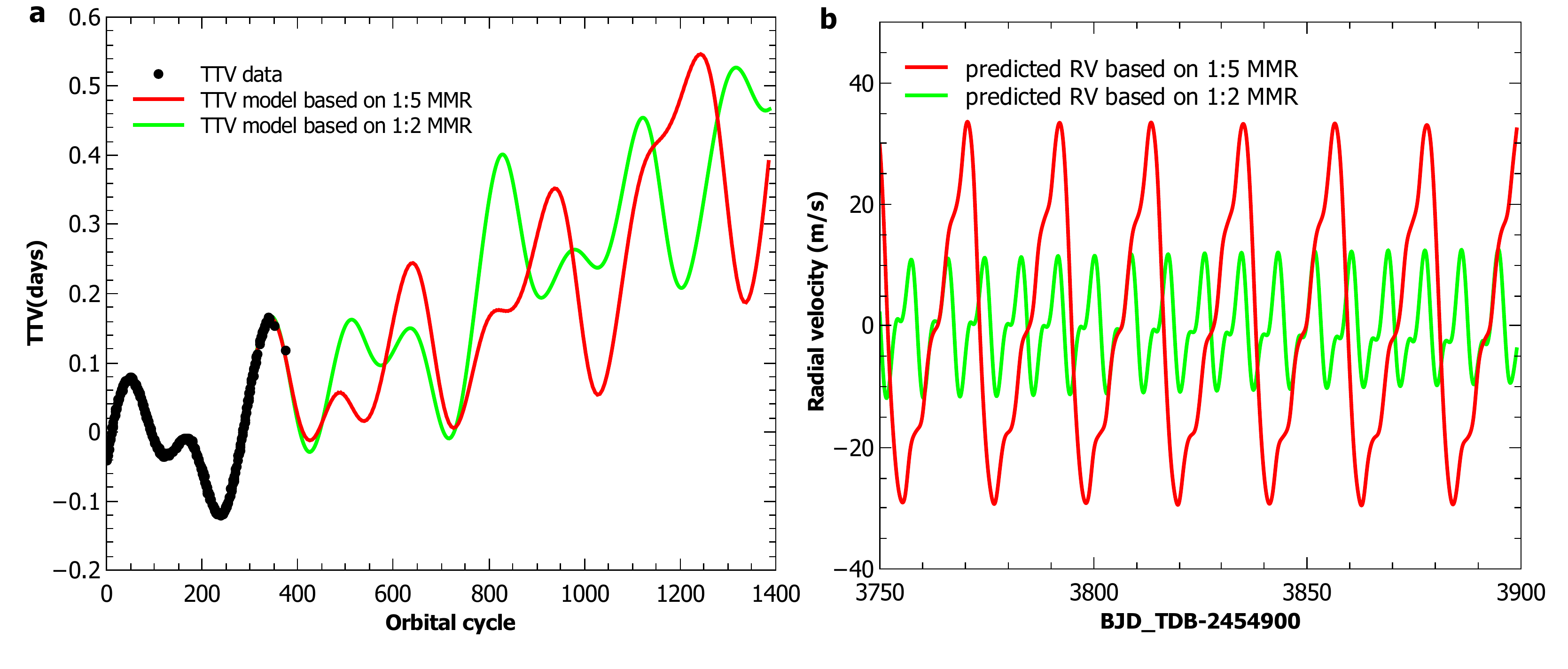}
\caption{Predicted TTV pattern of KOI-984$b$ and radial velocity curve of KOI-984 system in near future (red curves). \textbf{a}, the black dots represent measured TTV data; The last two black points are derived from the literature~\citep{Deck2015}, which are respectively observed on June 15, 2013 and September 26, 2013. \textbf{b}, the predicted RV curve is determined from the optimal TTV solution.}
  \label{plot7}
\end{figure*}

We plot the TTV and radial velocity curves for the near-future, as predicted by the optimal solution of joint analysis, in order to plan for follow-up observations (see Figure~\ref{plot7}). In Table~\ref{table4}, we list the predict transit timings, durations, impact parameters and radial velocity data based on 1:5 MMR. Note that we refine the instantaneous orbital period and mean anomaly of both planets at reference time $T_{BJD\_TDB}=2454957$ through replacing the fitting of TTVs with transit times. The ephemerids is used to calculate the TTV pattern of KOI-984$b$ (i.e., Figure~\ref{plot7} for 1:5 MMR) as follows:
\begin{equation}
T_{BJD\_TDB}=2454957.7837(14)+4.2881785(35)*N
\end{equation}
where $N$ is the orbital cycle number of KOI-984$b$.

\section{Dynamical stability}\label{dynamical stability}

\subsection{Stability analysis with the analytic criterion}

Both TTV inversion codes implement a Hill stability criterion that is suitable for two-planet systems~\citep{Gladman1993}. Although this initial check throws away the least stable systems, it cannot make it certain that the derived orbital architecture is stable in the long run, for example, like KOI-984 system with high mutual inclination and moderate eccentricity. Because such criterion is only valid under the conditions of low mutual inclinations ($I_{\rm mut} \leq 2\degr$)and small eccentricities ($e \le 0.2$). 

A long-term stability condition that is suitable to inclined, eccentric three-body system is provided by ~\cite{Mardling2001}:  
\begin{equation}
\frac{a_2}{a_1}> 2.8\left(1+\frac{m_3}{m_1+m_2}\right)^{2/5} \frac{(1+e_2)^{2/5}}{(1-e_2)^{6/5}}\left(1-\frac{0.3I_{mut}}{180\degr}\right) \quad .
\end{equation}
Here, $a_1$ and $a_2$ are the semi-major axes of the inner binary subsystem and the outer binary system (that is made of the outer body and the inner subsystem), respectively. $m_j$ denotes the mass of the $j$-th body ($j=1,2,3$), where $j=3$ is the outermost body. $e_2$ represents the orbital eccentricity of the outer binary system. $I_{\rm mut}$ is the mutual inclination of these two orbital planes.

This criterion is well satisfied by our orbital architecture derived from KOI-984$b$'s TTV inversion. Therefore, we conclude that the nominal orbital architecture of the KOI-984 system, which is extracted from inverting of measured KOI-984$b$'s TTVs, could meet the needs of long-term stability.

\subsection{Stability analysis using numerical $N$-body simulation}

After dynamical modeling of TTVs and TDVs, we obtain the initial orbital elements of the KOI-984 system at time $T_{BJD\_TDB}-2454900=57$. Basing on these initial conditions, we run an ensemble of $N$-body simulations to check the long-term stability of the system.

To meet our needs, we employ public $N$-body package REBOUND~\citep{Rein2012,Rein2015}, to evolve the system for 1~Myr. It should be noted that these integrations are only used to check whether the system remains stable for at least 1~Myr, rather than comprehensively study the dynamical evolution of the system. The stellar mass is fixed to 0.91~$M_\odot$, and the integration step is set to 0.05~times of the orbital period of the innermost planet. The initial semi-major axis of each planet is acquired from its period based on Kepler's third law. With the exception of orbital semi-major axis, the other parameters are fixed to the values derived from the TTV analysis, so as to improve the efficiency of the simulations. We sample a thousand initial values for the semi-major axis from a normal distribution $N(a_{0}, \sigma_{a})$, where $a_{0}$ is the nominal value of previous TTV analysis, and $\sigma_{a}$ is the uncertainty of the semi-major axis mainly from the large uncertainty of estimated host star's mass.

\begin{figure*}
\centering
\includegraphics[width=\linewidth]{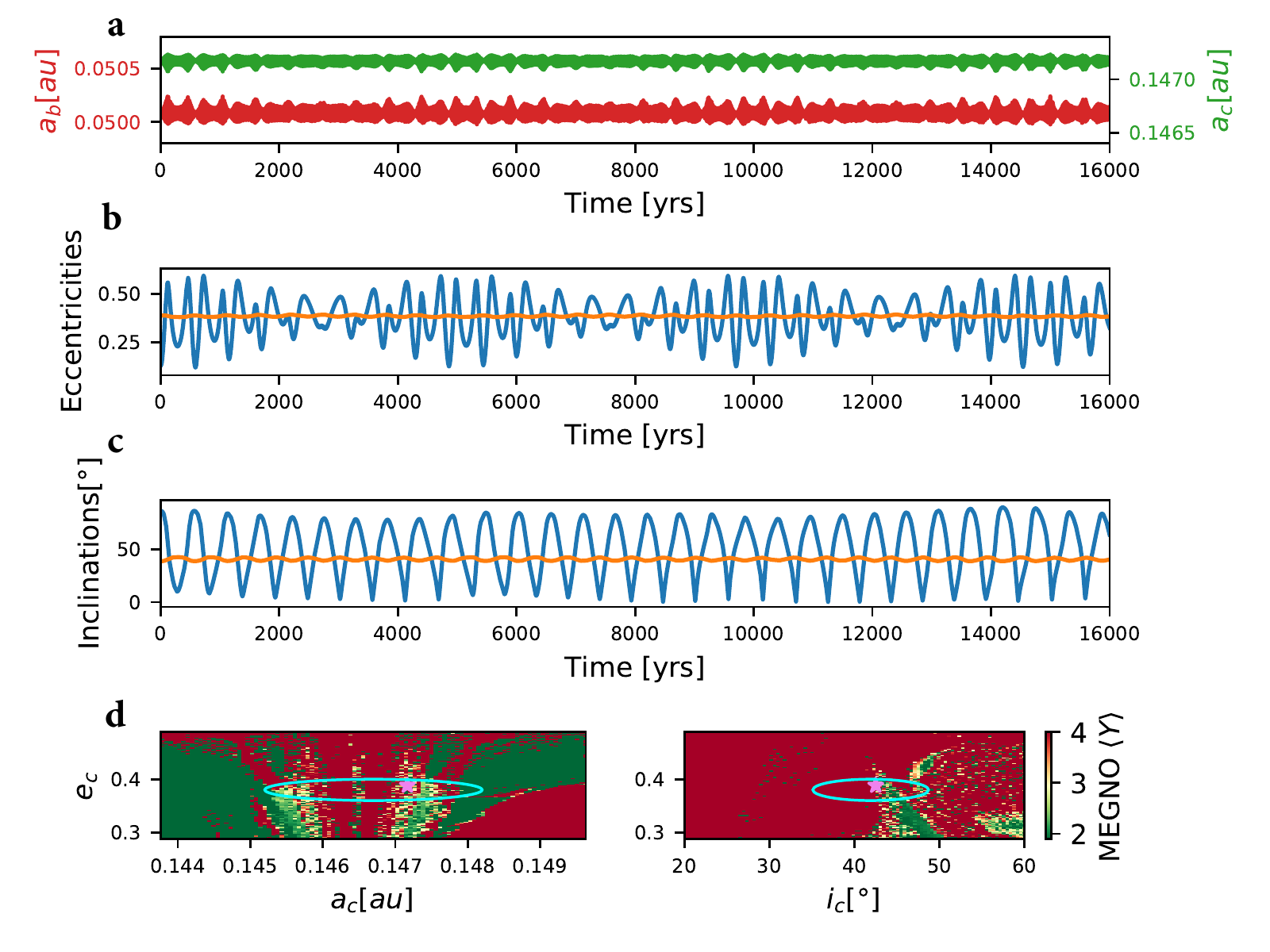}
\caption{The long-term evolution of KOI-984$b$ and KOI-984$c$'s orbital semi-axises ($a_b$, $a_c$), eccentricities ($e_b$, $e_c$) and inclinations ($i_b$, $i_c$), as well as MEGNO map of the system based on the orbital architecture near 1:5 MMR. \textbf{a}, the red and green curves denote $e_b$ and $e_c$, respectively. \textbf{b}, the blue and orange curves denote $a_b$ and $a_c$, respectively. The large oscillations in $e_b$ are induced by the outer highly-inclined warm-Jupiter KOI-984$c$ through the Lidov-Kozai mechanism. \textbf{c}, The blue and orange curves denote $i_b$ and $i_c$, respectively. The large oscillations in $i_b$ are also induced by KOI-984$c$ through the Lidov-Kozai mechanism. \textbf{d}, left panel: KOI-984 system's MEGNO map with various combinations between KOI-984$c$'s semi-major axis ($a_c$) and eccentricity ($e_c$). The regions with $\langle Y \rangle \le 2$ conservatively denote regular orbits (green shades), while the regions shaded in red denote the unstable orbits. Right panel: KOI-984 system's MEGNO map with various combinations between KOI-984$c$'s inclination ($i_c$) and eccentricity ($e_c$). As well, the associated posterior distributions within $1~\sigma$ confidence from TTV modeling are depicted approximately with two cyan ellipses in left and right panels, respectively. }
  \label{plot6}
\end{figure*}

The stable planetary systems are identified as those in which the minimum distance between the planets is never lower than the Hill stability criterion in the entire simulations. Systems that do violate this criterion generally appear to be short-lived due to close planet-planet encounters~\citep{Gladman1993}. We find that 7\% of all simulations could remain stable beyond 1~Myr. This relatively small fraction does not imply that the orbital architecture of KOI-984 derived from modeling TTVs and TDVs are less stable. Highly-inclined two-planet systems with moderate eccentricities are far more complicated than nearly coplanar systems~\citep{Naoz2016}. The long-term stability of highly-inclined systems could be ensured if the system was in a Lidov-Kozai (LK) resonant state. The LK resonant state offers a secular phase-protection mechanism for mutually inclined systems, even though the two orbits may suffer from large variations both in eccentricity and inclination (see Figure.~\ref{plot6} for further details about the large oscillations of both KOI-984$b$'s eccentricity and inclination). However, a rapid destabilization of highly mutually inclined systems is commonly observed, due to the chaotic region that develops around the stability islands of the LK resonance~\citep{Naoz2016,Volpi2019}. The long-term evolution of the KOI-984 system is probably determined by the combined effects of mean motion resonance and secular eccentric LK resonance, in addition to tidal and general relativistic effects. 

In order to determine whether the LK-resonant state is essential to ensure KOI-984 system's long-term stability, we have calculated the Mean Exponential Growth factor of Nearby Orbits (MEGNO) chaos indicator~\citep{Cincotta2000}. The orbits of KOI-984 system were numerically integrated with a large number of initial semi-major axes $a_c$, eccentricities $e_c$, and inclinations $i_c$ of planet $c$, but the other parameters of two planets were fixed to the nominal values obtained from the TTV inversion. We subsequently calculated the MEGNO maps on the different combinations of $a_c-e_c$ and $i_c-e_c$, respectively. We could not only identify the stable regions of the KOI-984 system on MEGNO maps, but can also visually obtain better constraints on orbital parameters from the requirement of long-term stability (See Figure.~\ref{plot6}).   

\section{Discussion} \label{discussion}

The tidal interaction between KOI984$b$ and its host star tends to spin up the host's rotational rate while transfer angular momentum from the planetary orbit to the host star. Over time its orbit will be circularised and the planet will likely spiral within the Roche limit of KOI-984$b$ and disintegrate with the aidding the exterior warm Jupiter KOI-984$c$. We compute the Roche limit as defined for a infinitely compressible object in Faber et al. (2005):
\begin{equation}
a_{Roche}=2.16R_p(M_\star/M_p)^{1/3}
\end{equation}
where $R_p$ and $M_p$ denote the planet radius and mass respectively and $M_s$ the stellar mass. According to the values listed in Table~\ref{table1}, the Roche limit for the planet KOI984$b$ is 0.01AU, which implys that KOI984$b$ is currently in the orbit that is much more than twice the Roche limit, so the pericenter distance of KOI-984$b$ will still larger than the Roche limit and the planet will not be immediately destroyed when it is on the high eccentricity phase of LK resonance (e.g. the maxima of KOI-984$b$'s eccentricities in Figure~\ref{plot6}).

\par The peculiar case in this system is the outer warm Jupiter KOI-984$c$, which has a moderate eccentricity and large mutual inclination, leading to a very compact (both planets' semi-major axes $a<0.15$~au), highly inclined warm-Jupiter system. Such kinds of systems are rare among the {\it Kepler} multi-planet systems, which mostly have low mutual inclinations and small eccentricities~\citep{Winn2015}. At present, only three multi-planet systems (i.e., Kepler-419, Kepler-108 and $\upsilon$~And) have well-measured large mutual inclinations~\citep{Dawson2014a,Mills2017,McArthur2010}, besides the KOI-984 system. Amongst the three previously known systems, the Kepler-108 is the most inclined, and its mutual inclination is smaller than 35\degr with a $1\sigma$ significance.

\par Such a peculiar system is not predicted by the current proposed formation models of warm Jupiters, and it will be a challenge for such models to explain the history of KOI-984. If KOI-984 system is the product of the high-eccentricity migration, it is expected to have experienced secular eccentricity oscillations with the aid of an outer close-by high-mass companion, and thus have high eccentricities ($e>0.4$) and no low-mass inner companions~\citep{Dong2014,Mustill2015,Mustill2017}; while if KOI-984$c$ forms in situ, it should have small eccentricities ($e<0.2$) and inner low-mass siblings with small mutual inclinations ~\citep{Huang2016,Petrovich2016,Boley2016}. Therefore, the KOI-984 system cannot be formed through either of these two pathways, even for a third pathway proposed for close-orbiting giant planets, namely disc migration, where planets exchange angular momentum with the proto-planetary disc and migrate to current low eccentricity orbits roughly aligned with the disc mid-plane ($I_{mtu}<$ 4\degr)~\citep{Kley2012,Baruteau2014}. 

Moreover, KOI-984$c$ is unlikely a proto-hot Jupiter that is currently undergoing high-eccentricity migration through tidal friction~\citep{Dawson2018}, since its orbital eccentricity is too small. Although high-eccentricity migration can be triggered by a nearby mutually-inclined massive body through secular gravitational interactions, causing the planetary eccentricity to undergo large oscillations~\citep{Petrovich2016}. For this to happen, however, a solar-mass perturber needs to be within a distance of $\sim 30$~au, or a Jupiter-mass perturber within $\sim 3$~au to overcome the general relativistic apsidal precession and then reach close enough at pericenter distance for effective tidal dissipation~\citep{Dong2014}. With a projected distance of about 600~au, KOI-984's closest stellar companion is too distant to excite a large eccentricity oscillation of KOI-984$c$.

We stress here that KOI-984 is one of several highly-inclined giant planetary systems with well-resolved 3D-orbital architecture. In addition, the dynamics of two planets of KOI-984 near the 5:1 MMR is probably dominated by a (high-order) mean motion resonance and/or the eccentric LK resonance~\citep{Naoz2016}, in addition to tidal and GR effects. The complexity of the KOI-984 system warrants further studies.

Recently, SuperWASP team reports the discovery of a similar compact, moderately eccentric warm-Jupiter system WASP-148 (i.e., $a_b=0.08$~au; $a_c=0.21$~au), which includes a transiting hot Jupiter and a moderatly eccentric, non-transiting warm Jupiter (i.e., $e_c\simeq 0.36$; $m_c\sin i_c\simeq 0.4 M_{\rm Jup}$), however, only a upper limit of both planets' mutual inclination is acquired through the stability analyses (i.e., $I_{\rm mut}< 35 \degr$)~\citep{Hebrard2020}. In addition, of total 260 Kepler planets and candidates that showed significant TTV signals, 121 are in single transiting systems~\citep{Holczer2016}; single transiting planets statistically have substantially larger eccentricities than multiple transiting planets~\citep{Xie2016}, which suggests that these single transiting systems with significant TTVs may have larger mutual inclinations than most {\it Kepler} multi-planet systems~\citep{Ida1993,Zhu2018}. Hence, such kinds of systems, like KOI-984 and WASP-148, enrich the knowledge on the diversity of extrasolar planetary systems; on the other hand, they inspire the reconsidering of warm-Jupiter population's formation and dynamical evolution.

%
\section{Conclusions} \label{conclusions}
\par In order to constrain the orbital properties of KOI-984$b$ and detect the unseen perturbing planet(s) that induce the huge TTVs of KOI-984$b$, we dynamically modeled KOI-984$b$'s TTVs and TDVs, assuming that either one, or two, additional perturbing planet(s) exist, respectively. The dynamical modeling of KOI-984 $b$'s TTVs and TDVs revealed the following results:
 \begin{enumerate} 
\item Both two-planet and three-planet models could well fit the TTVs of KOI-984$b$ from {\it Kepler} photometry. The optimal solution of three-planet model is actually a co-orbital configuration with an outer giant planet companion. However, the optimal orbital  architecture of the three-planet model cannot be stable for more than 1 Myr, so we believe that this configuration is less feasible for KOI-984 system.
\item The best two-planet model solution includes a $0.66 \pm 0.13 M_{\rm Jup}$ warm-Jupiter in a moderately eccentric orbit (i.e., called KOI-984$c$; $e_c=0.38\pm0.02$). This new planet is possible near 1:5 mean-motion resonance (MMR) with planet $b$ and its orbital plane mutually inclines by 45 $\pm$ 5\degr, relative to KOI-984$b$'s orbit. For another similar optimal solution, a Neptune-mass ($24 \pm 2\Mearth$) planet is near 1:2 mean-motion resonance (MMR) with planet $b$ and its orbital plane mutually inclines by 15 $\pm$ 5\degr.  
\item For the optimal solution of near 1:5 MMR, KOI-984 system cannot be predicted by the current proposed formation models of warm Jupiters.
\end{enumerate}

The planet KOI-984$c$ joins the small population of non-transiting exoplanets discovered using the TTV technique only (e.g., Kepler-46$c$~\citep{Nesvorny2012}, Kepler-419$c$~\citep{Dawson2014a}, Kepler-411$e$~\citep{Sun2019}, Kepler-82$f$~\citep{Freudenthal2019} and so on). Some of these, such as KOI-984$c$ (G$_{Gaia}$=12.4 mag), orbit magnetically-active host stars with fainter brightness compared to most other bright targets that are available to spectroscopic characterization with RV method and/or transmission spectroscopy. However, KOI-984 is an excellent target for CHEOPS~\citep{Broeg2013} to carry out follow-up observations. With longer baseline of transit observations, the mass and orbital architecture of KOI-984$c$ can be better constrained through dynamical modeling, which may shed light on the formation and dynamical history of this peculiar warm-Jupiter system. 

\begin{table}
\begin{center}
\caption{Basic data of the planet-host star KOI-984.}
\begin{tabular}{lrr} 
\hline
IDs:\\
KIC                   &        1161345  &   (1)  \\
TIC                   &      122784501  & (2)  \\
Gaia DR2                        &        2050249406656369920    & (3)  \\
\hline
RA (J2000)              &      19:24:11.6475     & (3) \\
DEC (J2000)             &   $+$36:50:22.8487      & (3)  \\       
$\mu$ RA (mas/yr)       &      $1.332  \pm 0.047$     & (3) \\
$\mu$ DEC(mas/yr)     &      $8.302  \pm 0.054$     & (3) \\
Parallax (mas)          &   $4.31 \pm 0.02 $     & (3)                        \\
Distance (pc)           &   $232.03 \pm 1.24$    & (3)    \\ 
\hline
Magnitudes:\\
$G$                             &     $12.4993 \pm 0.0010 $      & (3)          \\
\hline  
Stellar atmospheric parameters:\\
Effective temperature \teff\ [K] & $5295\pm150$         &(4)         \\
Surface gravity \logg\ [cm s$^{-2}$] & $4.54\pm0.15$ &(4)         \\
Metalicity abundance \met\ [dex] & $0.12 \pm 0.10$    &(4)         \\
\hline 
\multicolumn{3}{l}{ 
References: 
(1) Brown et al.~(\cite{Brown2011}); 
}
\\
\multicolumn{3}{l}{ 
 
(2) Stassun et al.~(\cite{Stassun2018a});  
}
\\
\multicolumn{3}{l}{ 
(3) Gaia Collaboration~(\cite{Gaia2018});
}
\\
\multicolumn{3}{l}{ 
(4) Deck et al.~(\cite{Deck2015}).}
\\
\end{tabular}
\label{table1}
\end{center}
\end{table}

\begin{table*}
\caption{Prior distributions for transit modeling and TTV inversion.}
\centering
\label{table2}
\begin{tabular}{llcc}
\hline\noalign{\smallskip}
&& \multicolumn{1}{c}{\bf KOI-984b} \\
\hline\noalign{\smallskip}
\emph{\hspace{0.3cm} Transit modeling parameters:} \\
Orbital period, $P$ & [d] & $4.288162712$ \\
\noalign{\smallskip}
Phase offset from the median data points, $(T_c-T_m)/P$ &  &$U(-0.1,0.1)$ \\
\noalign{\smallskip}
Ratio, $(R_p/R_*)^2$ &  & $U(0.005, 0.01)$ \\
\noalign{\smallskip}
Transit width, $T_{14}$ &[d]  & $U(0.01, 0.2)$ \\
\noalign{\smallskip}
Impact parameter, $b$ &  & $U(0.1, 1.5)$ \\
\noalign{\smallskip}
Stellar density, $\rho_*$ & [$\rho_\odot$]  & $N(1.526^{a}, 0.200)$ \\
\noalign{\smallskip}
\hline\noalign{\smallskip}
\emph{\hspace{0.3cm} Quadratic limb-dakerning coefficients$^{b}$:} \\
$q_1$  & &$0.481228$  \\
$q_2$  & &$0.338658$ \\
\noalign{\smallskip}
\hline\noalign{\smallskip}
\emph{\hspace{0.3cm} TTV inversion parameters:}  & & \textbf{KOI-984b}   &   \textbf{KOI-984c} \\
$P$  &[d] &$4.2881785$  &$U(1.0,65)$  \\
$M_p/M_*$  & &$0.00003$  &$U(10^{-8},0.009)$  \\
Inclination, $i$ & [\degr] & $86.7$ &U(26.7, 146.7)\\
$\sqrt{e} \cos{\omega}^c$  & &$U(-\sqrt{0.5},\sqrt{0.5})$  &$U(-1.0,1.0)$\\
$\sqrt{e} \sin{\omega}^c$  & &$U(-\sqrt{0.5},\sqrt{0.5})$  &$U(-1.0,1.0)$\\
$\Omega$ & [\degr] & $0$ &U(-60, 60)\\
$M_0$  &[\degr] &$U(0,360)$  &$U(0,360)$  \\
\noalign{\smallskip}
\hline\noalign{\smallskip}
\multicolumn{4}{l}{a: This value was obtained using the mass and radius emprical relation of~\citet{Torres2010} for single (post-) main-sequence stars.} \\
\multicolumn{4}{l}{b: using~\citet{Kipping2013}'s formula to implement the qudratic limb-darkening law.} \\
\multicolumn{4}{l}{$U(x_\text{min},  x_\text{max})$: uniform distribution between $x_\text{min}$ and $x_\text{max}$.} \\
\multicolumn{4}{l}{$N(\mu, \sigma)$: normal distribution with mean $\mu$ and standard deviation $\sigma$.} \\
\multicolumn{4}{l}{c: Here the constraint of $(\sqrt{e} \cos{\omega})^2+(\sqrt{e} \sin{\omega})^2 <=x_\text{min}^2$ is implemented in $\sqrt{e} \cos{\omega}$ and $\sqrt{e} \sin{\omega}$'s uniform sampling in $U(x_\text{min}, x_\text{max})$. } \\
\end{tabular}
\label{table2}
\end{table*}

\begin{table*}
\centering
\caption{KOI-984 system parameters.
The stellar parameters are obtained from Sect.~\ref{transit modeling}. The transit and orbital parameters are obtained from the transit modeling and TTV inversion based on 1:5 and 1:2 MMR models, respectly. (Sect.~\ref{TTV modeling}).
In particular, the mean ephemeris
of KOI-984$b$ are obtained from least square fitting to the transit timings of~\citet{Holczer2016}     
}            
\begin{minipage}{13.0cm} 
\renewcommand{\footnoterule}{}                          
\begin{tabular}{lccc}        
\hline           
\emph{\hspace{0.3cm}  \vspace{0.1cm} Stellar parameters:} 
 &      &                               \multicolumn{2}{c}{\textbf{KOI-984}}  \\
Stellar mass, $M_*$ &[\Msun] &                                                  \multicolumn{2}{c}{$ 0.928 \pm 0.031 $}  \\
Stellar radius, $R_*$ & [\Rsun] &                                       \multicolumn{2}{c}{$ 0.818 ^{+0.026}_{-0.025}$}   \\
Stellar density, $\rho_*$ & [$\rho_{\odot}$] &                                        \multicolumn{2}{c}{$ 1.70 \pm 0.15$}   \\
\hline
\emph{\hspace{0.3cm}\vspace{0.1cm} Transit and orbital parameters:}         & & \textbf{KOI-984b}   &   \\
Orbital period, $P$ & [d] &                             $4.2881785 \pm 0.0000035$    &      \\
Time of inferior conjunction, $T_c$ & $[\rm BJD_{\rm TDB}-2\,454\,900$] & $57.7837 \pm 0.0014$  &   \\
Ratio, $(R_p/R_{*})^2$ & &           $0.00229 \pm 0.00004$             &           \\
Transit duration, $T_{\rm 14}$ & [d] &                 $0.0660^{+0.0006}_{-0.0005}$            &      \\
Impact parameter, $b$ & &                               $0.83^{+0.01}_{-0.01}$                                &       \\
Normalized semimajor axis, $a/R_*$ &                   &                       $13.26 \pm 0.39$                                &        \\
Planet mass, $M_p$ & $[\Mearth]$  &       $17.3 \pm 3.0$\footnote{The planet mass is derived from the empirical relation of planet mass and radius\citep{Chen2018}.}                       &    \\
Planet radius, $R_p$  & $[\Rearth]$  &    $4.30 \pm 0.05$                       &   \\
Orbital semimajor axis, $a$  & [au] &          $0.0504 \pm 0.0006$& \\
Planet blackbody equilibrium temperature\footnote{Assuming a Bond 
albedo of 0 and a uniform heat redistribution to the night side.}, 
$T_{\rm eq}$    &       [K] &   $1022^{+31}_{-30}$    &        \\
\hline
\emph{\hspace{0.3cm}\vspace{0.1cm} \textbf{1:5 MMR}}         & &
\textbf{KOI-984b}   &   \textbf{KOI-984c}  \\
Orbital period, $P$ & [d] &                             $4.2881785 \pm 0.0000035$    &       $21.5120 \pm 0.0004$      \\
Planet mass, $M_p$ & $[\Mearth]$  &                  - -       &       $209.8 \pm 40.0$  \\
$\sqrt{e} \cos{\omega}$ & &                             $0.085 ^{+0.040}_{-0.084}$          &       $0.062 ^{+0.210}_{-0.089}$  \\ 
$\sqrt{e} \sin{\omega}$ & &                             $-0.336^{+0.036}_{-0.045}$          &       $0.613^{+0.013}_{-0.055}$  \\ 
Orbital eccentricity, $e$  &            &       $0.12 \pm 0.02$                       &       $0.38 \pm 0.02$      \\
Argument of periastron, $\omega$ & [\degr]        &       $-75 \pm 12 $            &       $84 \pm 18$         \\
Mutual inclination, $I_{mut}$ & [\degr]        &       - -            &       $45\pm 5.0$         \\
Mean anomaly at BJD$\,=2\,454\,957$, $M_0$ &  [\degr]     &       $309.9 ^{+32.5}_{-15.6}$ & $62.6^{+5.8}_{-2.6}$ \\
Orbital semimajor axis, $a$  & [au] &          $0.0504 \pm 0.0006$& $0.1467 \pm 0.0015$   \\
Planet blackbody equilibrium temperature, 
$T_{\rm eq}$    &       [K] &   $1022^{+31}_{-30}$    &       $597^{+52}_{-46}$ \\
\hline
\emph{\hspace{0.3cm}\vspace{0.1cm}  \textbf{1:2 MMR}}         & &
\textbf{KOI-984b}   &   \textbf{KOI-984c}  \\
Orbital period, $P$ & [d] &                             $4.2881785 \pm 0.0000035$    &       $8.5999 \pm 0.0004$      \\
Planet mass, $M_p$ & $[\Mearth]$  &       - - &       $24 \pm 2.0$  \\
$\sqrt{e} \cos{\omega}$ & &                             $0.057 ^{+0.011}_{-0.005}$          &       $-0.112 ^{+0.004}_{-0.011}$  \\ 
$\sqrt{e} \sin{\omega}$ & &                             $0.034^{+0.008}_{-0.006}$          &       $0.081^{+0.006}_{-0.011}$  \\ 
Orbital eccentricity, $e$  &            &       $0.006 \pm 0.002$                       &       $0.020 \pm 0.006$      \\
Argument of periastron, $\omega$ & [\degr]        &       $31 \pm 10 $            &       $147 \pm 12$         \\
Mutual inclination, $I_{mut}$ & [\degr]        &       - -            &       $15\pm 3.0$         \\
Mean anomaly at BJD$\,=2\,454\,957$, $M_0$ &  [\degr]     &       $23.7 ^{+6.2}_{-5.2}$ & $246.2^{+4.6}_{-3.9}$ \\
Orbital semimajor axis, $a$  & [au] &          $0.0504 \pm 0.0006$& $0.0796 \pm 0.0015$   \\
Planet blackbody equilibrium temperature,
$T_{\rm eq}$    &       [K] &   $1022^{+31}_{-30}$    &       $943^{+42}_{-39}$ \\
\hline       
\vspace{-0.5cm}
\end{tabular}
\end{minipage}
\label{table3} 
\end{table*}

\begin{table*}
\centering
\caption{Predicted transit timing, durations, impact parameters and radial velocity data by the optimal solution (1:5 MMR) of jointly analyzing the TTVs and TDVs.
}            
\begin{minipage}[t]{13.0cm} 
\renewcommand{\footnoterule}{}                          
\begin{tabular}{lccccccc}         
\hline\noalign{\smallskip}
& Orbital Cycle & Transit Timing($BJD_{TDB}$-2454900) & Impact Parameter & Transit Duration (d) & Epoch($BJD_{TDB}$-2454900) & RV (m/s) \\
\hline\noalign{\smallskip}      
& 0 & 57.7442 & 0.883 & 0.0676 & 3500 & -16.74 \\
& 1 & 62.0364 & 0.884 & 0.0675 & 3501 & -14.51 \\
& 2 & 66.3285 & 0.883 & 0.0675 & 3502 & -13.33 \\
& 3 & 70.6208 & 0.882 & 0.0678 & 3503 & -9.67 \\
& 4 & 74.9137 & 0.882 & 0.0678 & 3504 & -1.95 \\
& 5 & 79.2060 & 0.881 & 0.0679 & 3505 & 1.75 \\
& 6 & 83.4980 & 0.882 & 0.0678 & 3506 & 3.36 \\
& 7 & 87.7899 & 0.882 & 0.0678 & 3507 & 6.29 \\
& 8 & 92.0819 & 0.880 & 0.0681 & 3508 & 14.15 \\
& 9 & 96.3745 & 0.880 & 0.0681 & 3509 & 19.89 \\
& 10 & 100.6665 & 0.879 & 0.0682 & 3510 & 22.34 \\
& 11 & 104.9582 & 0.880 & 0.0681 & 3511 & 24.76 \\
& 12 & 109.2498 & 0.880 & 0.0681 & 3512 & 30.39 \\
& 13 & 113.5415 & 0.878 & 0.0684 & 3513 & 33.28 \\
& 14 & 117.8338 & 0.878 & 0.0685 & 3514 & 24.09 \\
& 15 & 122.1254 & 0.877 & 0.0686 & 3515 & 3.59 \\
& 16 & 126.4167 & 0.878 & 0.0685 & 3516 & -15.10 \\
& 17 & 130.7079 & 0.878 & 0.0685 & 3517 & -21.94 \\
& 18 & 134.9992 & 0.876 & 0.0687 & 3518 & -25.86 \\
& 19 & 139.2911 & 0.876 & 0.0688 & 3519 & -27.72 \\
& 20 & 143.5823 & 0.875 & 0.0690 & 3520 & -26.21 \\
& 21 & 147.8731 & 0.875 & 0.0689 & 3521 & -19.33 \\
& 22 & 152.1639 & 0.875 & 0.0689 & 3522 & -15.42 \\
& 23 & 156.4548 & 0.874 & 0.0691 & 3523 & -14.16 \\
& 24 & 160.7462 & 0.873 & 0.0692 & 3524 & -12.20 \\
& 25 & 165.0369 & 0.872 & 0.0694 & 3525 & -5.54 \\
& 26 & 169.3273 & 0.873 & 0.0693 & 3526 & 0.34 \\
& - - & - - & - - & - - & - - & - -  \\
& - - & - - & - - & - - & - - & - -  \\
& - - & - - & - - & - - & - - & - -  \\
& 1390 & 6018.7769 & 0.973 & 0.0469 &  &  \\
& 1391 & 6023.0721 & 0.973 & 0.0468 &  &  \\
& 1392 & 6027.3673 & 0.972 & 0.0470 &  &  \\
& 1393 & 6031.6629 & 0.979 & 0.0450 &  &  \\
& 1394 & 6035.9581 & 0.977 & 0.0456 &  &  \\
& 1395 & 6040.2535 & 0.978 & 0.0454 &  &  \\
& 1396 & 6044.5488 & 0.978 & 0.0453 &  &  \\
& 1397 & 6048.8442 & 0.977 & 0.0455 &  &  \\
& 1398 & 6053.1400 & 0.984 & 0.0434 &  &  \\
& 1399 & 6057.4352 & 0.982 & 0.0440 &  &  \\
& 1400 & 6061.7306 & 0.983 & 0.0438 &  &  \\
& 1401 & 6066.0260 & 0.983 & 0.0437 &  &  \\
& 1402 & 6070.3214 & 0.982 & 0.0439 &  &  \\
& 1403 & 6074.6172 & 0.989 & 0.0418 &  &  \\
& 1404 & 6078.9124 & 0.987 & 0.0424 &  &  \\
& 1405 & 6083.2078 & 0.988 & 0.0421 &  &  \\
& 1406 & 6087.5031 & 0.988 & 0.0421 &  &  \\
& 1407 & 6091.7984 & 0.987 & 0.0422 &  &  \\
& 1408 & 6096.0942 & 0.994 & 0.0401 &  &  \\
\hline       
\vspace{-0.5cm}
\end{tabular}
\end{minipage}
\label{table4} 
\end{table*}

\section*{Acknowledgements}

This paper includes data collected by the {\it Kepler} mission. Funding for the {\it Kepler} mission is provided by the NASA Science Mission directorate. We acknowledge the use of public TESS Alert data from pipelines at the TESS Science office and at the TESS Science Processing Operations Center. Resources supporting this work were provided by the NASA High-End Computing (HEC) Program through the NASA Advanced Supercomputing (NAS) Division at Ames Research Center for the production of the SPOC data products.

This research has made use of the NASA Exoplanet Archive, which is operated by the California Institute of Technology, under contract with the National Aeronautics and Space Administration under the Exoplanet Exploration Program. This research has made use of the VizieR catalogue access tool, CDS, Strasbourg, France. The original description of the VizieR service was published in A\&AS 143, 23. This work has made use of data from the European Space Agency (ESA) mission {\it Gaia} (\url{https://www.cosmos.esa.int/gaia}), processed by the {\it Gaia} Data Processing and Analysis Consortium (DPAC, \url{https://www.cosmos.esa.int/web/gaia/dpac/consortium}). Funding for the DPAC has been provided by national institutions, in particular the institutions participating in the {\it Gaia} Multilateral Agreement.

We are grateful to Eric Agol for providing valuable comments and suggestions that helped to improve this paper. L.S., S.G. and X.W. acknowledge financial support from National Natural Science Foundation of China (grants No. U1531121, No. 10873031, No.11473066 and No. 12003063). M.B.N.K. acknowledges support from the National Natural Science Foundation of China (grant 11573004). This research was supported by the Research Development Fund (grant RDF-16-01-16) of Xi'an Jiaotong-Liverpool University (XJTLU).

 This work has been in particular carried out under the frame between China Scholarship Council (CSC) and Deutscher Akademischer Austausch Dienst (DAAD). The joint research project between Yunnan Observatories and Hamburg Observatoy is funded by Sino-German Center for Research Promotion (GZ1419).
\section*{Data Availability}

{\it Kepler} data can be obtained  through the MAST archive \url{https://archive.stsci.edu/kepler}.\\

{\it TESS} data products can be accessed through the official NASA website \url{https://heasarc.gsfc.nasa.gov/docs/tess/data-access.html}.\\

The data that support the plots within this paper and other findings of this study are available from the corresponding authors upon reasonable request.



\bibliographystyle{mnras}
\bibliography{main}




\appendix

\newpage

\section*{Affiliations}
\noindent

\bsp	
\label{lastpage}
\end{document}